\documentclass[twocolumn,nodate,showpacs,preprintnumbers,nofootinbib,amsmath,amssymb,aps,prd]{revtex4}

\usepackage{graphicx}
\usepackage{amsmath,amssymb}

\def\b{$\bullet\,\,\,$}
\def\f{\frac}
\def\be{\begin{equation}}
\def\ee{\end{equation}}
\def\ba{\begin{eqnarray}}
\def\ea{\end{eqnarray}}

\def\F{\mathcal{F}}

\def\scri{\mathcal{I}}
\def\szm{\mathcal{I}^{o-}}

\def\szpr{\mathcal{I}^{o+}_{\rm R}}

\def\s1pr{\mathcal{I}^{1+}_{\rm R}}
\def\sbpr{{\mathcal{I}}^{+}_{\rm R}}
\def\spr{\mathcal{I}^{+}_{\rm R}}
\def\spl{\mathcal{I}^{+}_{\rm L}}
\def\h{\hat}

\def\ub{\underbar}

\def\b{\bar}
\def\Dp{\partial_{+}}
\def\Dm{\partial_{-}}
\def\dd{{\rm d}}

\def\k{\kappa}
\def\R{\mathbb{R}}
\def\y{{y}}

\newcommand{\ket}[1]{\ensuremath{|#1\rangle}}

\begin{document}

\title{Information is Not Lost in the Evaporation of 2-dimensional Black Holes}
    \author{Abhay Ashtekar${}^{1,2}$}
    \email{ashtekar@gravity.psu.edu}
    \author{Victor Taveras${}^1$}
     \email{victor@gravity.psu.edu}
    \author{Madhavan Varadarajan${}^{2,1}$}
    \email{madhavan@rri.res.in}
    \affiliation{${}^1$\!Institute for Gravitation and the Cosmos \&
     Physics Department, Penn State, University Park, PA 16802,
    USA\\
    ${}^2$\!Raman Research Institute, Bangalore, 560 080 India}

\begin{abstract}

We analyze Hawking evaporation of the
Callan-Giddings-Harvey-Strominger (CGHS) black holes from a
quantum geometry perspective and show that information is not
lost, primarily because the quantum space-time is sufficiently
larger than the classical. Using suitable approximations to
extract physics from quantum space-times we establish that:
i)future null infinity of the quantum space-time is sufficiently
long for the the past vacuum to evolve to a pure state in the
future; ii) this state has a finite norm in the future Fock space;
and iii) all the information comes out at future infinity; there
are no remnants.
\end{abstract}


\maketitle

In his celebrated paper \cite{swh1}, Hawking showed that in
quantum field theory on a fixed black hole space-time the vacuum
state at past null infinity $\scri^-$ evolves to a thermal state
on $\scri^+$. Thus, in this external field approximation, pure
states evolve into mixed; information is lost. Hawking also drew a
candidate Penrose diagram including back reaction and suggested
that information loss would persist. There has since been a large
body of literature on the issue using diverse methods, models and
approximations. More recently, the AdS/CFT conjecture has been
used to argue that information cannot be lost. However, this
reasoning requires a negative cosmological constant and even in
that case a \emph{space-time} description of the evaporation
process is still lacking.

In this Letter we analyze the issue of information loss using the
1+1 dimensional CGHS model \cite{cghs}. The model is well suited
because it shares most of the conceptual complications of realistic
4-dimensional black holes but is technically simpler to analyze.
Therefore it drew a great deal of attention in the early nineties
(see, e.g., \cite{reviews} for excellent reviews). Although a firm
conclusion could not be reached due to limitations of semi-classical
methods that were used, partial results suggested to many authors
that information is probably lost.

Our analysis is motivated by the fact that quantum geometry leads
to resolution of space-like singularities in a number of simple
models (see, e.g., \cite{sing}). This resolution provides an
entirely new perspective on the problem \cite{ab2}. For, much of
the older discussion assumed, as Hawking did, that the future
boundary of the relevant space-time consists not just of $\scri^+$
but also a piece of the initial classical singularity (See FIG.
1). Since part of the `in' state falls into the singularity, it is
not surprising that the `out' state at $\scri^+$ fails to capture
the full information contained in the `in' state at $\scri^-$. By
contrast, if the singularity is resolved, this potential sink of
information is removed. We will argue that in the quantum
extension of the classical CGHS space-time, $\scri^+$ is long
enough to register all the information contained in the `in'
state. Although our considerations are motivated by loop quantum
gravity,
in this Letter we will use the more familiar Fock quantization
since the main argument is rather general.
%
\begin{figure}
\begin{center}
\includegraphics[width=2.2in,height=2in,angle=0]{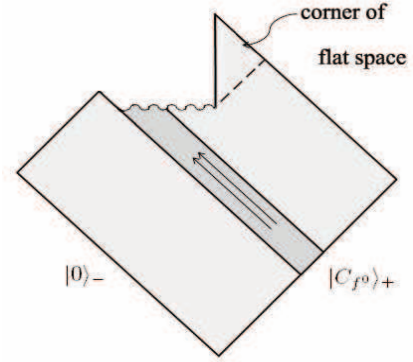}
\caption{A Penrose diagram of an evaporating CGHS black hole,
motivated by \cite{swh1}. Information can be lost in the singularity
represented by the wiggly line.} \label{Oldpicture}
\end{center}
\end{figure}

\emph{Classical Theory:} Fundamental fields of the CGHS model are
the space-time metric $g$, a dilaton $\phi$ and a massless scalar
field $f$. The action is given by
\ba S(g,\phi,f) = \f{1}{G}\textstyle{\int}\!\! &\dd^2V&\!\!
e^{-2\phi}\, \left(R + 4 g^{ab}\nabla_a\phi \nabla_b \phi + 4
\k^2\right)
\nonumber\\
&-& \f{1}{2} \textstyle{\int} \dd^2V\, g^{ab}\nabla_a f \nabla_b f
\ea
where $R$ is the scalar curvature of $g$ and $\k$ is a constant
(with dimensions of inverse length). Let $M_o \equiv \R^2$ and fix
on it a Minkowski metric $\eta$. Denote by $\scri^{o\pm}$ its null
infinity. We will be interested in physical metrics $g$ which
approach $\eta$ at $\szm$. Denote by $z^\pm$ the advanced and
retarded null coordinates of $\eta$ so that $\eta_{ab} =\ -
\partial_{(a}z^+ \, \partial_{b)}z^-$ and set $\partial_\pm
= \partial/\partial z^{\pm}$. Finally, set
\be\label{defs}  \Phi = e^{-2\phi}\quad {\rm and}\quad g^{ab} =
\Theta^{-1} \Phi\, \eta^{ab} \equiv \Omega\, \eta^{ab}\, . \ee
Our fundamental fields will be $\Phi, \Theta, f$. They satisfy:
\ba \label{de1} \Box_{(g)}\, f = 0 &\Leftrightarrow&
\Box_{(\eta)}f = 0\nonumber\\
\Dp\,\Dm\, \Phi + \k^2 \Theta &=& G\, T_{+-}\nonumber\\
\Phi \Dp\,\Dm \ln \Theta &=& -G\, T_{+-} \ea
and
\ba \label{bcs} -\Dp^2\, \Phi + \Dp\,\Phi \Dp\, \ln \Theta &=& G
T_{++}\nonumber\\
-\Dm^2\, \Phi + \Dm\,\Phi \Dm\, \ln \Theta &=& G T_{- -}\ea
where $T_{+ -}, T_{+ +}, T_{- -}$ are the $z^\pm$ components of
the stress energy tensor of $f$. If (\ref{bcs}) are imposed at
$\szm$, they are propagated by (\ref{de1}). Therefore we will
refer to (\ref{de1}) as \emph{dynamical equations} and ensure that
(\ref{bcs}) are satisfied by choosing appropriate boundary
conditions at $\szm$. In the classical theory $T_{-+}$ vanishes
identically but in quantum theory it is non-zero because of the
trace anomaly.

Because $f$ satisfies the wave equation on $(M_o, \eta)$, it can
be naturally decomposed into left and right moving modes
$f_\pm(z^\pm)$. In the sector of the theory of interest to us,
$f_- =0$ and a black hole forms because of the gravitational
collapse of $f_+$ (FIG. 1). To express the solution explicitly, it
is simplest to use coordinates $x^\pm$:
%
\be \k x^+ = e^{\k z^+}, \quad {\rm and} \quad \k x^- = -e^{-\k
z^-}\, .\ee
Then, for any given $f_+$, the classical solution satisfying the
appropriate boundary conditions at $\szm$ is given by
\cite{class}:
\ba \label{sol1} \Theta (z^\pm) &=& - \k^2 x^+\,x^-\nonumber\\
\Phi(z^\pm) &=& \Theta (z^\pm) - \f{G}{2}\textstyle{\int_0^{x^+}}
\dd\bar{x}^+\, \textstyle{\int_0^{\bar{x}^+}} \dd
\bar{\bar{x}}^+\, (\partial
f_+/\partial \bar{\bar{x}}^+)^2  \nonumber\\
&-& \f{G}{2} \textstyle{\int_{0}^{{x}^-}} \dd {\bar{x}}^-\,
\textstyle{\int_{0}^{\bar{x}^-}} \dd\bar{\bar{x}}^-\, (\partial
f_-/\partial \bar{\bar{x}}^-)^2\, . \ea
This brings out the fact that the true degree of freedom lies just
in the matter field $f$; the geometry and the dilaton is
determined algebraically from $f$. (The term containing $f_-$
vanishes classically but is important for quantum considerations
that follow.)

The solution is regular on all of $M_o$. How can there be a
singularity and a black hole then? To answer this question let us
examine the physical metric $g^{ab} = \Omega \eta^{ab} \equiv
\Theta^{-1} \Phi\, \eta^{ab}$. Now, although $\Omega$ (and hence
$g^{ab}$) is a well defined tensor field on all of $M_o$, $\Phi$
vanishes on a space-like line. Along this line $g^{ab}$ also
vanishes and its curvature becomes infinite. Thus $\Phi =0$ is the
singularity of the \emph{physical} metric $g$. Is this a black
hole singularity? Right null infinity $\spr$ of $g$ is a proper
subset of $\szpr$ (of $\eta$) \cite{reviews}. However detailed
analysis shows that it \emph{is} complete with respect to $g$ and
its past does not contain the singularity. Thus the singularity
\emph{is} hidden behind a horizon with respect to $\spr$. However,
left null infinity $\spl$ is incomplete to the future. So,
strictly we cannot conclude that we have a black hole with respect
to $\spl$ \cite{gh}. Fortunately, $\spl$ does not play a direct
role in the analysis of Hawking effect and information loss.


\emph{Quantum Theory:}
Consider the space of all classical solutions. If $f\not=0$, the
manifold $M_{(g)}$ on which the physical metric $g$ is well
defined is a proper subset of $M_o$, which however varies from
solution to solution. Therefore, the appropriate arena is the
manifold $M_o$ defined by the fiducial $\eta$.
This suggests that we represent $\hat{f}_\pm$ as an operator
valued distribution on the Fock space $\F_+ \otimes \F_-$
associated with $(M_o,\eta)$ and define $\h\Theta$ and $\h\Phi$
also on this Hilbert space. Since $f_-=0$ classically, the quantum
sector of interest is spanned by states $\Psi$ of the type
$\ket{C_{f^o}}_+ \otimes \ket{0}_-$ on $\szm$, where $f^o$ is any
suitably regular profile of $f_+$ and $C_{f^o}$ the coherent state
in $\F_+$ peaked at $f^o$. The span of these states is $\F_+
\otimes \ket{0}_-$.

We will use the Heisenberg picture. The operator $\h{g}^{ab} =
\h{\Omega}\eta^{ab}$ will define the quantum geometry on $M_o$. The
basic operators $\h{f} =\h{f}_+ + \h{f}_-, \h\Theta, \h\Phi$ must
satisfy the operator version of dynamical equations (\ref{de1}) and
appropriate boundary conditions at $\szm$. More precisely, detailed
considerations imply that a mathematical
quantum theory of the model would result if we can:\\
i) Solve (\ref{de1}) for operators $\h{f}, \h\Theta, \h\Phi$, where
$T_{+-}$ is replaced by the trace anomaly $T_{+-}(\h{g})$ defined by
the conformal factor $\h\Omega$;
and,\\
ii) Ensure that
%
%
\emph{at} $\szm$, $\h\Theta$ and $\h\Phi$ are given by the operator
versions of (\ref{sol1}), with $(\partial f_\pm/\partial
\bar{\bar{x}}^\pm)^2$ replaced by $:(\partial \h{f}_\pm/\partial
\bar{\bar{x}}^\pm)^2:$, where the normal ordering is defined by
$\eta$. (Operator versions of (\ref{bcs}) are then automatically
satisfied at $\szm$.)\\
It is likely that this framework can be made fully rigorous along
the lines of the D\"utsch and Fredenhagen \cite{df} approach to
interacting fields in Minkowski space-time.

The key physical questions are: i) In the solution, are $\h\Theta$
and $\hat\Phi$ well-defined everywhere on $M_o$?; ii) Does the
operator valued distribution $\h\Omega$ vanish anywhere? If it
did, the quantum metric $\h{g}^{ab} = \h\Omega\,\eta^{ab}$ could
be singular there; and, iii) What is the \emph{physical}
interpretation of the Heisenberg state in the quantum geometry of
$\h{g}^{ab}$? The third question is crucial for extracting physics
from the mathematical framework. While proposals of formulating
the quantum theory in terms of operators have appeared in the
literature (see, e.g. \cite{miko}), to our knowledge our specific
formulation is new and the third question in particular had not
received due attention. In the rest of the Letter we will
introduce two approximation schemes to answer these questions.
These schemes will also shed light on the exact framework.

\emph{Bootstrapping:} Although the quantum versions of the
dynamical equations (\ref{de1}) form a closed hyperbolic system
for $\h\Theta$ and $\h\Phi$, they are difficult to solve exactly.
To develop intuition for the quantum geometry that would result,
it is instructive to simplify this task by a bootstrapping
procedure. Begin with a seed metric $\h{g}_o$ and use it to
calculate the trace anomaly $\h{T}_{+-}$, feed the result in the
right side of the quantum dynamical equations, solve them, and
denote the solution by $\h\Theta_{1}, \h\Phi_{1}$ and
$\h{g}_{1}^{ab}$. In the second step, use $\h{g}_{1}^{ab}$ as the
seed metric and continue the cycle in the hope of obtaining better
and better approximations to the closed system of interest.

Let us begin by choosing $\h{g}_o$=$\eta$. Then, the first cycle
can be completed. The solution on all of $M_o$ is $\h{\Theta}_1= -
\k^2 x^+ x^-$ and $\h{\Phi}_1 = \h{\Theta}_1 -
\f{G}{2}\textstyle{\int_0^{x^+}} \dd\bar{x}^+\,
\textstyle{\int_0^{\bar{x}^+}} \dd \bar{\bar{x}}^+\\ :\!(\partial
\h{f}_+/\partial \bar{\bar{x}}^+)^2\!:\,  - \f{G}{2}\,
\textstyle{\int_{0}^{{x}^-}} \dd {\bar{x}}^-\,
\textstyle{\int_{0}^{\bar{x}^-}} \dd\bar{\bar{x}}^-\, :\!(\partial
\h{f}_-/\partial \bar{\bar{x}}^-)^2\!:$ where normal ordering is
defined by $\eta$. How does this truncated solution fare with
respect to the key physical questions? $\h{\Theta}_1$ happens to
be a c-number and $\h{\Phi}_1$ can be shown to be an operator
valued distribution in a well-defined sense. They are regular
everywhere on $M_o$ whence \emph{the quantum geometry determined
by $\h{g}_1^{ab}$ is also regular on all of $M_o$} already at the
first approximation. The expectation values $\Phi_1 := \langle
\h{\Phi}_1 \rangle$ and $g_1^{ab} := \langle \h{g}_1^{ab}\rangle$
turn out to reproduce just the classical solution $\Phi_{\rm
class}$ and $g^{ab}_{\rm class}$ given by (\ref{sol1}). In
particular, $g_1^{ab}$ vanishes along a space-like line and
\emph{its} Ricci scalar diverges there. However, one can also
calculate the fluctuations of $\h{g}_1^{ab}$ (after suitable
smearings since it is an operator valued distribution) and they
are very large near that line. Therefore, the expectation value is
a poor representation of quantum physics which is perfectly
regular there.

The answer to the third physical question is even more
interesting. We know that the quantum state of $\h{f}_-$ is simply
the vacuum state $\ket{0}_-$ on $(M_o,\eta)$. The question is:
What is its physical interpretation on the space-time $(M_1,g_1)$
that results at the end of the first cycle? Following \cite{swh1},
one can carry out detailed analysis at late times.
There are again two conceptual elements: i) Since $y_1^-$ defined
by the asymptotic time translation on $g_1$ is non-trivially
related to $z^-$, there is positive and negative frequency mixing
between modes of $\hat{f}_-$ defined using $z^-$ and those defined
using $y_1^-$; and, ii) Since $g_1^{ab}=g^{ab}_{\rm class}$, its
right future null infinity $\s1pr$ is a proper subset of $\szpr$
of $\eta^{ab}$. Hence one has to trace over modes in $\szpr -
\s1pr$. The result is that for the algebra of observables in
$(M_1, g_1)$,\, $\ket{0}_-$ reduces to the density matrix
$\h{\rho}_1 := ({\rm const}) \,\, \exp -\beta \h{H}$, where $\beta
= 2\pi/\hbar \k$, and $\h{H}$ the Hamiltonian of $\h{f}_-$ at
$\s1pr$. Thus, at this order one recovers the Hawking effect.

To summarize, the regular quantum geometry of $\h{g}_1$ does not
define some exotic sector of the theory, but has the right
physical content. Since $\h\Theta_1,\,\h\Phi_1,\,\h{g}_1$ is an
exact solution to the truncated version of full quantum equations,
it provides useful intuition for the nature of quantum geometry in
the full theory. The next step in the bootstrapping is to start
the second cycle using $\h{g}_1$ as the seed metric.
Unfortunately, the resulting quantum equations are now almost as
difficult to solve as the exact ones. There is however another
approximation that is well suited for analyzing the issue of
information loss, which we now introduce.

\emph{Mean Field Approximation (MFA):} Rather than using a seed
metric, let us return to the closed system of exact quantum
dynamical equations, take their expectation values, and solve the
resulting equations in the mean field approximation, i.e., by
replacing expectation values of the type $\langle F(\h\Theta,
\h\Phi)\rangle$ by $F(\langle \h\Theta\rangle, \langle
\Phi\rangle)$. Viability of this approximation requires a large
number $N$ of matter fields so that quantum fluctuations
$\h\Theta$ and $\h\Phi$ can be neglected relative to those in the
matter fields. This large $N$ approximation has been examined in
some detail in the literature \cite{largeN,reviews} and initial
data near $\szm$ have been evolved numerically. Examination of
marginally trapped surfaces in the resulting solutions shows that
the Bondi mass at right null infinity of the mean field metric
steadily decreases (essentially) to zero due to quantum radiation.
This was often taken to mean that one can attach to the
numerically evolved space-time a \emph{ `corner' of flat space} as
in Hawking's original guess (see FIG.1). However, a definitive
statement could not be made because, even when $N$ is large,
fluctuations of geometry become dominant in the space-time
interior making MFA invalid there.

Our new observation is that the \emph{key to the information loss
issue lies in the geometry near future infinity and MFA should be
valid there.} Thus, we will assume that: i) the exact quantum
equations can be solved and the expectation value $\bar{g}^{ab}$
of $\h{g}^{ab}$ admits a smooth right null infinity $\sbpr$ which
coincides with $\szpr$ \emph{in the distant past} (i.e. near
$i^o_{\rm R}$);\, ii) MFA holds in a neighborhood of $\sbpr$; and,
\, iii) Flux of quantum radiation vanishes at some finite value of
the affine parameter ${\y}^-$ of $\sbpr$ defined by the asymptotic
time translation of $\b{g}$. All three assumptions were standard
in previous analyses. Indeed, one cannot even meaningfully ask if
information is lost unless the first two hold.

A priori, $\sbpr$ may be only a proper subset of $\szpr$ and no
assumption is made about $i^+$ of $\b{g}$. However, existence of
$\sbpr$ implies that as we go to large $z^+$ values along constant
$z^-$ lines, $\b\Phi := \langle \h{\Phi}\rangle$ and $\b\Theta :=
\langle\h{\Theta}\rangle$ admit asymptotic expansions of the form:
\ba {\b\Phi} &=& A(z^-) e^{\k z^+} + B(z^-) +
O(e^{-\k z^+})\nonumber\\
{\b\Theta} &=& \ub{A}(z^-) e^{\k z^+} + \ub{B}(z^-) + O(e^{-\k
z^+})\, .\ea
The MFA equations determine $\ub{A}$ and $\ub{B}$ in terms of $A$
and $B$. Furthermore, $\y^-$ adapted to the asymptotic time
translation of $\b{g}$ is given by\, $\k\exp -\k\y^- =A$.
Finally, the MFA equations imply that there is a balance law at
$\sbpr$:
\ba \label{balance}\f{d}{\dd\y^-}\big[ \f{\dd{B}}{\dd\y^-} &+& \k
{B}\, +\, \f{N \hbar G}{24}\, \big(\f{\dd^2\y^-}{\dd z^{-2}}\,
(\f{\dd\y^-}{\dd z^-})^{-2}\,
\big)\,\big] \nonumber\\
&=& - \f{N \hbar G}{48}\, \big[\f{\dd^2\y^-}{\dd z^{-2}}\,
(\f{\dd\y^-}{\dd z^-})^{-2}\,\,\big]^2 \ea
It is natural to identify the quantity in square brackets on the
left side as $G m_{\rm B}$, where $m_{\rm B}$ the Bondi mass, and
the right side as the energy flux at $\sbpr$. These definitions have
the desired properties that the energy flux is positive definite and
$m_{\rm B}$ vanishes in flat space (which is an MFA solution).
The first two terms in the expression of $m_{\rm B}$ yield
Hayward's formula \cite{sh} of Bondi mass in the classical theory;
the third term is a quantum correction.

A key question now is: How large is $\sbpr$ compared to $\szpr$?
By assumption they coincide in the distant past near $i^o_{\rm
R}$. One can show that $\y^- = C z^- + D$ (with $C,D$ constants)
on the entire future region of $\sbpr$ where the quantum flux
vanishes. 
Hence $\sbpr = \szpr$\, (see FIG. 2). This implies that to
interpret $\ket{0}_-$ at $\sbpr$ we no longer have to trace over
any modes; in contrast to the situation encountered in our
bootstrapping discussion, all modes of $\h{f}_-$ are now
accessible to the asymptotically stationary observers of $\b{g}$.
The vacuum state $\ket{0}_-$ of $\eta$ is pure also with respect
to $\b{g}$.

\begin{figure}
\begin{center}
\includegraphics[width=2.2in,height=2in,angle=0]{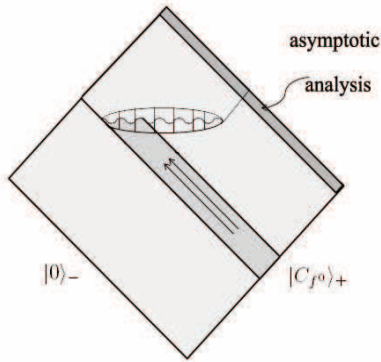}
\caption{Proposed Penrose diagram. The mean field approximation is
used in the shaded region in near $\sbpr$. Quantum fluctuations of
geometry are large in the interior region around the wiggly line
representing the putative classical singularity.}
\label{Newpicture}
\end{center}
\end{figure}

But is it in the asymptotic Fock space of $\b{g}$? Calculation of
Bogoluibov coefficients shows that the answer is in the
affirmative because $\y^- = Cz^- + D$ in the future and boundary
conditions imply that $\y^-$ approaches $z^-$ exponentially
quickly in the distant past. Thus, the interpretation of
$\ket{0}_-$ with respect to $\b{g}$ is that it is a pure state
populated by pairs of particles at $\sbpr$. \emph{There is neither
information loss nor remnants} whose quantum state is correlated
with the state at $\sbpr$. 

\emph{Summary:}\, A key simplification in the CGHS model is that
the matter field satisfies just the wave equation on
$(M_o,\eta^{ab})$. Therefore, given initial data on $\szm$, we
already know the state everywhere both in the classical and the
quantum theory. However, the state derives its physical
interpretation from geometry which is a complicated functional of
the matter field. We do not yet know the quantum geometry
everywhere. But already at the end of the first cycle of
bootstrapping we found that $\h{g}^{ab}_1$ is well-defined (and
nowhere vanishing) everywhere on $M_o$. So it seems reasonable to
assume that the full $\h{g}^{ab}$ would also be singularity-free.
To pose questions about information loss, one has to assume that
its expectation value $\b{g}$ admits future right null infinity
$\sbpr$ which, a priori, could may be only a portion of $\szpr$ of
$\eta$. But then the MFA equations imply that $\sbpr$ in fact
coincides with $\szpr$ and the exact quantum state $\ket{0}_-$ is
a pure state in the asymptotic Fock space of $\b{g}^{ab}$. The
S-matrix is unitary and there is no information loss. The Penrose
diagram (FIG. 2) we are led to is significantly different from
that based on Hawking's original proposal (FIG. 1). In particular,
the quantum space-time does not end at a future singularity and is
larger than that in FIG. 1. The singularity is replaced by a
genuinely quantum region and, in contrast to an assumption that
was often made, space-time need not be flat to its `future'.
Finally, although $\hat{g}^{ab} = \hat\Omega\, \eta^{ab}$,
$\hat\Omega$ is an operator and is not required to be positive
definite. In the region around the wiggly line of FIG. 2, quantum
fluctuations of $\hat\Omega$ are large and of either sign (where
the negative sign corresponds to interchanging time-like and
space-like directions). Thus, the global causal structure is not
that of Minkowski
space-time. 

We emphasize however that a full solution to the quantum equations
is still lacking. This is needed to prove the validity of our
assumptions and to calculate, everywhere on $\sbpr$ the function
$\y^-(z^-)$ that determines the detailed physical content of
$\ket{0}_-$ at $\sbpr$. Nonetheless, using what we already know,
we can answer the oft raised question: When does the `information'
come out? Following the standard strategy, let us use a basis at
$\sbpr$ analogous to that of \cite{swh1}, trace over modes to the
future of the point where the Bondi mass vanishes and ask if the
resulting state is approximately pure. In our framework the answer
is in the affirmative. Thus, most of the `information' comes out
with the quantum radiation. This issue as well as several others
that have been raised in the literature will be discussed in the
detailed paper.

\textbf{Acknowledgment:} We would like to thank Klaus Fredenhagen,
Steve Giddings, Jim Hartle, Don Marolf and Samir Mathur for
stimulating discussions. This work was supported in part by the
NSF grant PHY-0456913 and the Eberly research funds of Penn State.

\end{document}